\documentclass[preprint,aps]{revtex4}
\usepackage{amsmath}
\usepackage{amssymb}

\input{tcilatex}

\begin{document}

\title{Multispinon excitations in the spin $S=1/2$ antiferromagnetic
Heisenberg model}
\author{Yu-Liang Liu}
\affiliation{Department of Physics, Renmin University, Beijing 100872, \\
People's Republic of China}

\begin{abstract}
With the commutation relations of the spin operators, we first write out the
equations of motion of the spin susceptibility and related correlation
functions that have a hierarchical structure, then under the "soft cut-off"
approximation, we give a set of equations of motion of spin susceptibilities
for a spin $S=1/2$ antiferromagnetic Heisenberg model, that is independent
of whether or not the system has a long range order in the low
energy/temperature limit. Applying for a chain, a square lattice and a
honeycomb lattice, respectively, we obtain the upper and the lowest
boundaries of the low-lying excitations by solving this set of equations.
For a chain, the upper and the lowest boundaries of the low-lying
excitations are the same as that of the exact ones obtained by the Bethe
ansatz, where the elementary excitations are the spinon pairs. For a square
lattice, the spin wave excitation (magnons) resides in the region close to
the lowest boundary of the low-lying excitations, and the multispinon
excitations take place in the high energy region close to the upper boundary
of the low-lying excitations. For a honeycomb lattice, we have one kind of
"mode" of the low-lying excitation. The present results obey the
Lieb-Schultz-Mattis theorem, and they are also consistent with recent
neutron scattering observations and numerical simulations for a square
lattice.
\end{abstract}

\maketitle

\section{\protect\bigskip Introduction}

Spins are neither bosons nor fermions, and their commutation relations make
spin problems so difficult. A spin system is a subject in which there are
few exactly solvable models which are nontrivial. Only a few of the models
have solutions\cite{1,1a,2,2a} which are well understood, in spite of the
fact that many of them have been intensely studied\cite{3,3ab,3a}. The most
challenge of a spin system is that there is absent of an analytical method
directly applied for it beyond one dimension (1D) without the help of the
slave particle representations.

Instead of directly studying a spin system, one usually maps it to a
many-body problem which is a strongly interacting system. For example, a
spin $S=1/2$ chain may be mapped exactly into an interacting spinless
fermion system, with the help of the Jordan-Wigner transformation\cite{4,4a}
which is believed to be valid only for one dimension. While for a
high-dimensional spin system, the spin operators are usually represented by
slave bosons/fermions with some constraint conditions, then the system is
mapped into an interacting boson/fermion system which has still not been
very successful because of the strong interactions among bosons/fermions\cite%
{6,7}. In contrast with a bosonic/fermionic system in which the basic
low-lying excitation are quasiparticles obeying Bose-Einstein/Fermi-Dirac
statistics, the low-lying excitations\cite{8} of a spin $S=1/2$
antiferromagnetic system are magnons with spin $S=1$, where a magnon may be
seen as a bound state (triplet) of two spinons that each spinon has a spin $%
S=1/2$. However, in the 1D case, the spinons are nearly deconfined, and they
become the elementary low-lying excitations\cite{9,10} of the system. The
calculations\cite{12a,12b,12c,12d,12e} based on the exact solution of the
Bethe ansatz and the recent neutron scattering measurements on
one-dimensional\cite{11} spin $S=1/2$ Heisenberg antiferromagnets strongly
support the picture that the spinons are the elementary low-lying
excitations. Recent neutron scattering experiment\cite{12} shows that for
two-dimensional (2D) spin $S=1/2$ Heisenberg antiferromagnets, the spinons
may be nearly deconfined in some short wave-length regions of the Brillouin
zone (BZ).

Theoretically, the 2D Heisenberg antiferromagnet has been extensively
studied by a variety of numerical approaches that try to completely
understanding of these experimental observations with nearly deconfined
spinons\cite{12,13} or multi-magnon excitation\cite{14,15,16}, where there
does not have a convincing unambiguous evidence to support which one of them
as the low-lying excitations in short wave length region of the BZ. However,
the numerical calculations based on the exact solution of the Bethe ansatz
and the neutron scattering experimental observations unambiguously show that
the low-lying excitations of the spin $S=1/2$ systems are distributed a broad
region in the frequency and momentum plane with the upper and the lowest
boundaries, especially in the 1D case that the lowest boundary mainly
represent the two spinon excitations, that has the same dispersion as that
of usual spin wave, while the upper boundary represents the multispinon
(pairs) excitation, that has a dispersion with a period twice that of the
spin wave.

In contrast to usual slave particle methods, the equation of motion of
Green's function approach can be a good candidate in studying of the
low-lying excitations of a spin $S=1/2$ antiferromagnetic system, in which
one can easily write out the equations of motion of all high order
correlation functions appearing in the equation of motion of Green's
function. The equations of motion of Green's function and related high order
correlation functions are tightly coupled with each other, and they have a
simply hierarchic structure that is an unclosed set of equations. In each
level of this hierarchic structure there are many correlation functions
where they construct a subset of equations.

In Refs.[17,18], the authors used the equation of motion of Green's function
to have calculated the low-lying excitation of a 1D spin $1/2$ Heisenberg
model with usual cut-off approximations taken for high order correlation
functions, and they had obtained the low-lying excitation spectrum that is
consistent with the exact one\cite{8} only for small momentum. At larger
momentum, however, their results heavily deviate from the exact ones.
Moreover, it is hard to have the spin wave (the lowest boundary) of the
low-lying excitations as that ones by the Bethe ansatz and the recent
neutron scattering measurements, because in the previous calculations of
Refs.[17,18] they cannot self-consistently calculated equations of motion of
the high order correlation functions appearing in the same level of this
hierarchic structure.

In this paper, we first write out of a complete hierarchic structure of the
equations of motion of multiple-point correlation functions. Instead of
taking usual cut-off approximations for high order correlation functions, we
solve self-consistently the equations of motion of multiple-point
correlation functions under "soft cut-off " approximations, then we can
obtain the upper and lowest boundaries of the low-lying excitations of the
magnons/(pair) spinons in the whole BZ for a spin $S=1/2$ antiferromagnetic
Heisenberg model in 1D and 2D. For 1D, the upper and lowest boundaries of
the low-lying excitations have the same dispersion as that ones of the Bethe
ansatz in the whole BZ, and for 2D, they are completely consistent with the
recent neutron scattering experimental observations and numerical
simulations.

This paper is organized as follows. In section II, we give a detail
explanation of our present method. Under the "soft cut-off" approximations%
\cite{19}, we write out the general expressions of equations of motion of
the transverse and longitudinal spin susceptibilities in the $N=1$ and $N=2$
levels, respectively, in Section III. Then we apply these equations of
motion of the transverse and longitudinal spin susceptibilities for the 1D
and 2D cases, and calculate the low-lying excitation spectrums of the
magnons/(pair) spinons in the whole BZ in Sections IV-VI. Finally we give
our conclusions and discussions in Section VII. More technical calculations
for the high order multiple-point correlation functions are put in the
Appendixes.

\section{Basic idea of the algebraic equation of motion approach}

For the spin operators of the spin $S=1/2$, they have a SU(2) symmetry. As
an unperturbable theory, we extend the hierarchical Green's function approach%
\cite{19} to spin $S=1/2$ magnetic systems, called algebraic equation of
motion approach. The basic idea of the algebraic equation of motion approach
is that: if we calculate the equation of motion of the correlation function
of an operator $\widehat{A}\left( t\right) $, that is written out in the
Heisenberg representation, we need to calculate the commutation relation of
the operator $\widehat{A}$ with the Hamiltonian $H$, $\left[ \widehat{A},H%
\right] $, that may produce a new operator $\widehat{B}$, then we calculate
again the commutation relation of the operator $\widehat{B}$ with the
Hamiltonian, $\left[ \widehat{B},H\right] $, that may produce another new
operator $\widehat{C}$, in turn we calculate again the commutation relation
of the operator $\widehat{C}$ with the Hamiltonian, $\left[ \widehat{C},H%
\right] $, and so on. Finally, we have a finite number of these operators
that are elementary ingredients as in writing out of the EOMs of the
correlation function of the operator $\widehat{A}\left( t\right) $ and
related multiple-point correlation functions that are defined by these new
operators.

As applying this approach for the spin $S=1/2$ antiferromagnetic Heisenberg
model, we use the algebraic commutation relations of spin operators with the
Hamiltonian of the system to write out the equations of motion (EOMs) of the
spin susceptibility and related multiple-point correlation functions, \ and
these EOMs of the spin susceptibility and the related multiple-point
correlation functions have a hierarchic structure denoted by a level
parameter $N$ (see appendix A). The EOMs of the multiple-point correlation
functions belonging to the same $N$ level construct a subset of equations,
in which there emerge some other multiple-point correlation functions
belonging to the $N+1$ level, like that for electronic systems.

For the spin $S=1/2$, the spin operators $\widehat{\mathbf{s}}_{i}$ satisfy
the relations, $\left( \widehat{s}_{i}^{z}\right) ^{2}=\frac{1}{4}$, and $%
\widehat{s}_{i}^{+}\widehat{s}_{i}^{-}=\frac{1}{2}+\widehat{s}_{i}^{z}$.
With these relations, the EOMs of the related multiple-point correlation
functions can be significantly simplified. For example, in the EOM of a
related multiple-point correlation function belonging to the $N$-level,
under the above relations of the spin operators there emerge some
multiple-point correlation functions belonging to the $N-1$ level, as a
simple approximation (called "soft cut-off" approximation\cite{19}), we can
discard those multiple-point correlation functions belonging to the $N+1$
level to make the set of equations of the multiple-point correlation
functions be closed. Based on this prominent character of spin $S=1/2$
system, we can effectively calculate the low-lying excitation spectrums of
the spins under the approximation to only keeping the related multiple-point
correlation functions belonging to the $N=2$ level and discarding all other
high order ones.

In the following sections, we use the bipartite sublattice representation to
write out the EOMs of spin susceptibility and related multiple-point
correlation functions, and all calculations about the EOMs of the
multiple-point correlation functions are made on the lattice sites. The
prominent advantage of the bipartite sublattice representation is that it
can greatly simplify our calculating for the high order multiple-point
correlation functions that are in fact the tensors whose indexes denoted by
the lattice site coordinates. Finally, we only retain the results that are
independent of the bipartite sublattice representation.

\section{Basic equations of motion of the spin susceptibility}

If we only consider the contributions of the related multiple-point
correlation functions belonging to $N=1$ level, under the "soft cut-off "
approximation\cite{19}, we can obtain the following equations of motion
(EOMs) of the transverse and longitudinal spin susceptibilities,%
\begin{equation}
\left[ \omega ^{2}-\Delta _{0}^{zz}\right] \chi _{0iq}^{zz}(\omega )=A\delta
_{iq}-\frac{1}{2}\sum_{j}\left( J_{ij}^{\bot }\right) ^{2}\widetilde{\chi }%
_{0jq}^{zz}(\omega )  \label{4a}
\end{equation}%
\begin{equation}
\left[ \omega ^{2}-\Delta _{0}^{zz}\right] \widetilde{\chi }%
_{0iq}^{zz}(\omega )=A_{iq}-\frac{1}{2}\sum_{j}\left( J_{ij}^{\bot }\right)
^{2}\chi _{0jq}^{zz}(\omega )  \label{4aa}
\end{equation}%
\begin{equation}
\left[ \omega ^{2}-\Delta _{0}^{+-}\right] \widetilde{\chi }%
_{0iq}^{-+}(\omega )=-C_{iq}-\frac{1}{2}\sum_{j}J_{ij}^{\bot }J_{ij}^{z}\chi
_{0jq}^{-+}(\omega )  \label{4b}
\end{equation}%
\begin{equation}
\left[ \omega ^{2}-\Delta _{0}^{+-}\right] \chi _{0iq}^{-+}(\omega )=B\delta
_{iq}-\frac{1}{2}\sum_{j}J_{ij}^{\bot }J_{ij}^{z}\widetilde{\chi }%
_{0jq}^{-+}(\omega )  \label{4c}
\end{equation}%
where $\Delta _{0}^{zz}=\frac{1}{2}\sum_{j}\left( J_{ij}^{\bot }\right) ^{2}$%
, $\Delta _{0}^{+-}=\frac{1}{4}\sum_{j}\left[ \left( J_{ij}^{\bot }\right)
^{2}+\left( J_{ij}^{z}\right) ^{2}\right] $, $A=\frac{1}{2}%
\sum_{j}J_{ij}^{\bot }<\widehat{X}_{ij}^{\left( +\right) }>$, $A_{iq}=\frac{1%
}{2}\sum_{j}J_{ij}^{\bot }<\widehat{X}_{ji}^{\left( +\right) }>\delta _{jq}$%
, $B=2\omega <\widehat{s}_{i}^{z}>+2\sum_{j}J_{ij}^{z}<\widehat{s}_{i}^{z}%
\widehat{\tau }_{j}^{z}>+\sum_{j}J_{ij}^{\bot }<\widehat{s}_{i}^{+}\widehat{%
\tau }_{j}^{-}>$, and $C_{iq}=2\sum_{j}J_{ij}^{\bot }<\widehat{s}_{j}^{z}%
\widehat{\tau }_{i}^{z}>\delta _{jq}+\sum_{j}J_{ij}^{z}<\widehat{s}_{j}^{+}%
\widehat{\tau }_{i}^{-}>\delta _{jq}$, where $\widehat{s}_{i}^{\pm }=%
\widehat{s}_{i}^{x}\pm i\widehat{s}_{i}^{y}$. These EOMs of the spin
susceptibility are universal for a general spin $S=1/2$ antiferromagnetic
Heisenberg model, and they can be used to calculate its low-lying excitation
spectrum on a variety of lattice sites, where the static constants can be
self-consistently determined by a set of equations of equal-time spin
susceptibility derived from the relation, $\widehat{s}_{i}^{+}\widehat{s}%
_{i}^{-}=\frac{1}{2}+\widehat{s}_{i}^{z}$, and sum rules. However, under
this simple approximation, the above EOMs of the spin susceptibility can
only give the reliable upper boundary of the low-lying excitation and they
cannot give usual spin wave excitation that survive in the large momentum
and low energy limit region. In order to studying the lowest low-lying
excitations residing in the large momentum region, we have to consider the
contributions of the high order related multiple-point correlation functions
belonging to $N=2$ level.

Under the "soft cut-off" approximation, as including the contributions of
the related multiple-point correlation functions belonging to the $N=2$
level (see Appendix C), for example, we can obtain the following EOMs of the
transverse spin susceptibility (for simplicity, taking $J^{\bot }=J^{z}=J$),%
\begin{eqnarray}
\left[ \omega ^{2}-\Delta \left( \omega \right) \right] \chi
_{iq}^{-+}(\omega ) &=&B\delta _{iq}-\sum_{j}J_{ij}^{U}\widetilde{\chi }%
_{jq}^{-+}(\omega )  \notag \\
&&-\sum_{jl}J_{ij}\Gamma _{lij}\left( \omega \right) \left[ \chi
_{iq}^{-+}(\omega )-\chi _{lq}^{-+}(\omega )\right]  \label{4a1}
\end{eqnarray}%
\begin{eqnarray}
\left[ \omega ^{2}-\Delta \left( \omega \right) \right] \widetilde{\chi }%
_{iq}^{-+}(\omega ) &=&-C_{iq}-\sum_{j}J_{ij}^{U}\chi _{jq}^{-+}(\omega )
\notag \\
&&-\sum_{jl}J_{ij}\Gamma _{lij}\left( \omega \right) \left[ \widetilde{\chi }%
_{iq}^{-+}(\omega )-\widetilde{\chi }_{lq}^{-+}(\omega )\right]  \label{4a2}
\end{eqnarray}%
where $B=2\omega <\widehat{s}_{i}^{z}>+\sum_{j}J_{ij}\left[ 2<\widehat{s}%
_{i}^{z}\widehat{\tau }_{j}^{z}>+<\widehat{s}_{i}^{+}\widehat{\tau }_{j}^{-}>%
\right] $, $C_{iq}=\sum_{j}J_{ij}\left[ 2<\widehat{s}_{i}^{z}\widehat{\tau }%
_{j}^{z}>+<\widehat{s}_{i}^{+}\widehat{\tau }_{j}^{-}>\right] \delta _{jq}$,
$\Delta \left( \omega \right) =\sum_{j}J_{ij}^{U}\left( \omega \right) $,
and $J_{ij}^{U}\left( \omega \right) =\frac{1}{2}\left( J_{ij}\right)
^{2}+J_{ij}\sum_{l}\left[ \Pi _{lij}\left( \omega \right) +\Pi _{lji}\left(
\omega \right) \right] $. The coefficients $\Gamma _{lij}\left( \omega
\right) $ and $\Pi _{lij}\left( \omega \right) $ can be approximately
written as that,%
\begin{equation}
\Gamma _{lij}\left( \omega \right) =\frac{J_{ij}\left( J_{jl}\right)
^{2}\left( 1-\delta _{il}\right) }{16D_{zz}\left( \omega \right) }\left( 1-%
\frac{3J^{2}}{4}\frac{D_{\tau z}\left( \omega \right) }{D_{X}\left( \omega
\right) }+\frac{J^{2}}{2D_{\tau z}\left( \omega \right) }\right)  \label{4d1}
\end{equation}%
\begin{eqnarray}
\Pi _{lij}\left( \omega \right) &=&2\Gamma _{lij}\left( \omega \right) +%
\frac{J_{ij}\left( J_{jl}\right) ^{2}\left( 1-\delta _{il}\right) }{%
16D_{\tau z}\left( \omega \right) }\left( 1+\frac{J^{2}}{2D_{zz}\left(
\omega \right) }\right)  \notag \\
&&+\frac{D_{\tau z}\left( \omega \right) }{2D_{X}\left( \omega \right) }%
\frac{3J_{ij}\left( J_{il}\right) ^{2}\left( 1-\delta _{jl}\right) }{8}
\label{4d2}
\end{eqnarray}%
where $D_{\tau z}\left( \omega \right) =\omega ^{2}-\frac{J^{2}}{2}$, $%
D_{zz}\left( \omega \right) =\omega ^{2}-J^{2}$, $D_{X}\left( \omega \right)
=\left( \omega ^{2}-J^{2}\right) ^{2}-\frac{J^{4}}{4}$. Obviously, as $%
\omega >J$, the coefficients $\Gamma _{lij}\left( \omega \right) $ and $\Pi
_{lij}\left( \omega \right) $ are positive, and in the limit, $\omega
/J\rightarrow \infty $, they go to zero, then the Eqs.(\ref{4a1},\ref{4a2})
are reduced to the Eqs.(\ref{4b},\ref{4c}), which means that the latter is
the high energy limit of the former. To the contrary, in the limit, $\omega
/J\rightarrow 0$, the coefficient $J_{ij}^{U}\left( \omega \right) $ goes to
zero, and the coefficient $\Gamma _{lij}\left( \omega \right) $ becomes a
constant, $-J_{ij}J_{jl}\left( 1-\delta _{il}\right) /32J$. In this case, we
can obtain the lowest boundary of the low-lying excitations of the spins.
The poles appearing in the coefficients $\Gamma _{lij}\left( \omega \right) $
and $\Pi _{lij}\left( \omega \right) $ are artificial, that originate from
the approximations taken for the high order related multiple-point
correlation functions belonging to the $N=2$ level. The summation over the
site variables of $\Gamma _{lij}\left( \omega \right) $ and $\Pi
_{lij}\left( \omega \right) $ is very clear for a chain, while for the high
dimensional lattice case, such as for a square lattice and a honeycomb
lattice, this summation must be careful, due to the number of the next
nearest neighbor sites becomes large. For a square lattice, the summation
over the next nearest neighbor sites of the site $x_{i}$ is restricted as
the sites $x_{i\pm 2\mathbf{e}_{x}}$ and $x_{i\pm 2\mathbf{e}_{y}}$, where $%
\mathbf{e}_{x}$ and $\mathbf{e}_{y}$ are the x-axis and y-axis unit vectors,
respectively, and we discard other sites, such as, $x_{i\pm \mathbf{e}%
_{x}\pm \mathbf{e}_{y}}$ and $x_{i\pm \mathbf{e}_{x}\mp \mathbf{e}_{y}}$,
due to these sites can be reached from the site $x_{i}$ by two different
ways.

In comparison with the EOMs of the transverse spin susceptibility in Eqs.(%
\ref{4b},\ref{4c}), the ones in Eqs.(\ref{4a1},\ref{4a2}) have a prominent
character that there emerges the $\Gamma _{lij}\left( \omega \right) $ term
on the right hand side, which is contributed by the high order
multiple-point correlation functions. In the bipartite sublattice
representation, the $\Gamma _{lij}\left( \omega \right) $ term represents
the relation between the spin susceptibility on different sites of the same
spin ingredient, which is survived in the low energy limit. While, the $%
J_{ij}^{U}\left( \omega \right) $ term describes the relation between the
spin susceptibility on the nearest neighbor sites of the different spin
ingredients, which is going to zero in the low energy limit. Under the
condition of the locally short range antiferromagnetic correlation, the
former one corresponds to the effect of twice spin-flipping process on
different sites, which is a pair of kink and anti-kink in 1D, and the latter
one is the effect of one spin-flipping process on the nearest neighbor sites,
which is a kink in 1D. Based on these considerations, it is convincible to
assume that the $\Gamma _{lij}\left( \omega \right) $ term represents the
low-lying excitations of magnons (pairs of spinons), and the $%
J_{ij}^{U}\left( \omega \right) $ term describes the low-lying excitations
of nearly deconfined spinons. While, the magnons and nearly deconfined
spinons are coexisting in the mid energy range where $\Gamma _{lij}\left(
\omega \right) $ and $J_{ij}^{U}\left( \omega \right) $ are finite. This
picture is completely consistent with the exact one by the Bethe ansatz\cite%
{9} in 1D.

\section{A spin chain}

For simplicity, we first consider a spin chain without the longitudinal
coupling $J_{ij}^{z}=0$, (XY model) to calculate the low-lying excitations
of spins. According to the Eqs.(\ref{4a},\ref{4aa}), we have the low-lying
excitation spectrum of the XY model,%
\begin{equation}
\varepsilon _{0k}^{XY}=\sqrt{2}J^{\bot }|\sin \left( \frac{k}{2}\right) |
\label{5}
\end{equation}%
where choosing the lattice constant one. Based on the exact solution of the
Bethe ansatz of the spin $S=1/2$ chain XY model\cite{9a}, the authors\cite%
{10} had calculated the low-lying spectrum which is that, $\varepsilon
_{k}^{Bethe}=2J^{\bot }|\sin \left( \frac{k}{2}\right) |$. The spectrums $%
\varepsilon _{0k}^{XY}$ and $\varepsilon _{k}^{Bethe}$ both have the same
dispersion in the range of momentum, $-\pi \leq k\leq \pi $, and the
difference between them is only their coefficients. The $\varepsilon
_{0k}^{XY}$ is the upper boundary of spectrum of the XY model\cite{8}, and
it represents the low-lying excitations of nearly deconfined spinons\cite{9}%
. In order to have the lowest boundary of spectrum, like that in Eqs.(\ref%
{4a1},\ref{4a2}), it needs to calculate the contributions coming from the
high order multiple-point correlation functions to the spin susceptibility $%
\chi _{iq}^{zz}(\omega )$ and $\widetilde{\chi }_{iq}^{zz}(\omega )$. After
including the contributions of the multiple-point correlation functions
belonging to $N=2$ level (see Appendix B), we obtain the lowest boundary of
spectrum of the XY model,%
\begin{equation}
\varepsilon _{k}^{XY}=J^{\bot }|\sin \left( k\right) |  \label{6}
\end{equation}%
which is the low-lying excitations (spin wave) of magnons, and it possesses
double periodicity of $|\sin \left( k\right) |$, like the exact one\cite%
{8,9,10}.

In the case of the isotropic couplings $J^{\perp }=J^{z}=J$, with the Eqs.(%
\ref{4a}-\ref{4c}), we can obtain the following low-lying excitation
spectrums,%
\begin{eqnarray}
\varepsilon _{0k}^{L} &=&\sqrt{2}J|\cos \left( \frac{k}{2}\right) |  \notag
\\
\varepsilon _{0k}^{U} &=&\sqrt{2}J|\sin \left( \frac{k}{2}\right) |
\label{7}
\end{eqnarray}%
Here $\varepsilon _{0k}^{L/U}$ represent the lowest- and upper-boundary of
the low-lying excitations only considering the contributions of the high
order related multiple-point correlation functions belonging to $N=1$ level.
Notice that the low-lying excitation spectrum $\varepsilon _{0k}^{L}$ will
disappear without using the bipartite sublattice representation, thus it may
be an artificial result produced by the bipartite sublattice representation.

The spectrum $\varepsilon _{0k}^{U}$ has the same dispersion on the whole
range of $k$ as that of the $S=1$ low-lying excitation of the Bethe ansatz, $%
\varepsilon _{Uk}^{Bethe}=\pi J|\sin \left( \frac{k}{2}\right) |$, while the
spectrum $\varepsilon _{0k}^{L}$ is different from another one of the $S=1$
low-lying excitation spectrum of the Bethe ansatz, $\varepsilon
_{Lk}^{Bethe}=\frac{\pi }{2}J|\sin \left( k\right) |$. After including the
contributions of the multiple-point correlation functions belonging to $N=2$
level, the momentum dependence in $0\leq k\leq \pi $\ of the upper boundary
of the low-lying excitation spectrum is the same as that of $\varepsilon
_{0k}^{U}$, only their coefficients are modified, which describes the
low-lying excitations of nearly deconfined spinons; However, the lowest
boundary of the low-lying excitations is heavily modified, and it possesses
double periodicity of $|\sin \left( k\right) |$, like that for the XY model,
and it has the same momentum dependence in $0\leq k\leq \pi $ as the exact
one $\varepsilon _{Lk}^{Bethe}$ (see below).

With the help of the Eqs.(\ref{4a1},\ref{4a2}) that included the
contribution of the high order related multiple-point correlation functions
belonging to $N=2$ level, we obtain the following equations that can be used
to determine the low-lying excitation spectrums of the system,%
\begin{equation}
\omega ^{2}=\Delta \left( \omega \right) \left[ 1-\cos \left( k\right) %
\right] -\Gamma \left( \omega \right) \left[ 1-\cos \left( 2k\right) \right]
\label{8a}
\end{equation}%
\begin{equation}
\omega ^{2}=\Delta \left( \omega \right) \left[ 1+\cos \left( k\right) %
\right] -\Gamma \left( \omega \right) \left[ 1-\cos \left( 2k\right) \right]
\label{8b}
\end{equation}%
where $\Delta \left( \omega \right) =J^{2}+4\Gamma \left( \omega \right) +%
\frac{J^{4}}{4D_{\tau z}\left( \omega \right) }\left( 1+\frac{J^{2}}{%
2D_{zz}\left( \omega \right) }\right) +\frac{3J^{4}D_{\tau z}\left( \omega
\right) }{4D_{X}\left( \omega \right) }$ and $\Gamma \left( \omega \right) =%
\frac{J^{4}}{8D_{zz}\left( \omega \right) }\left( 1-\frac{3J^{2}}{4}\frac{%
D_{\tau z}\left( \omega \right) }{D_{X}\left( \omega \right) }+\frac{J^{2}}{%
2D_{\tau z}\left( \omega \right) }\right) $. Here the Eq.(\ref{8b}) will
disappear without using the bipartite sublattice representation. Thus, the
physical low-lying excitations of the system is determined by the Eq.(\ref%
{8a}) which is independent of the bipartite sublattice representation. The
momentum dependence of the low-lying excitations is controlled by the
factors, $1-\cos \left( k\right) $ and $1-\cos \left( 2k\right) $, that show
different behavior around $k\sim 0$ and $k\sim \pi $. In the regime around $%
k\sim 0$, the factors $1-\cos \left( k\right) $ and $1-\cos \left( 2k\right)
$ both go zero, and the possible excitation region of the spins is narrow.
However, in the momentum regime around $k\sim \pi $, there emerges a broad
region of the low-lying excitation in the frequency and momentum plane,
where its upper and lowest boundaries can be determined by the Eq.(\ref{8a}%
), since the factor $1-\cos \left( k\right) $ goes to $2$, while the factors
$1-\cos \left( 2k\right) $ goes to zero.

In the high energy limit, $\omega /J\rightarrow \infty $, the coefficient $%
\Delta \left( \omega \right) $ takes the value, $\Delta \left( \omega
\right) =J^{2}$, and the coefficient $\Gamma \left( \omega \right) $ goes to
zero, $\Gamma \left( \omega \right) =0$, in which we can have the upper
boundary of the low-lying excitation. In the low energy limit, $\omega
/J\rightarrow 0$, the coefficient $\Delta \left( \omega \right) $ goes to
zero, $\Delta \left( 0\right) =0$, and the coefficient $\Gamma \left( \omega
\right) $ becomes a constant, $\Gamma \left( 0\right) =-J^{2}/16$, where the
lowest boundary of the low-lying excitations can be determined. Under these
two limits, we have the lowest and the upper boundaries of the low-lying
excitations,%
\begin{eqnarray}
\varepsilon _{k}^{L} &=&\frac{\sqrt{2}}{4}J|\sin \left( k\right) |  \notag \\
\varepsilon _{k}^{U} &=&\sqrt{2}J|\sin \left( \frac{k}{2}\right) |  \label{9}
\end{eqnarray}%
Comparing with the Eq.(\ref{7}), we find that the high order multiple-point
correlation functions belonging to $N=2$ level determines the lowest
boundary $\varepsilon _{k}^{L}$ of the low-lying excitation (spin wave) of
the magnons, and they have little influence on the upper boundary $%
\varepsilon _{k}^{U}$ of the low-lying excitations of the nearly deconfined
spinons. In the whole region of the momentum, $0<k<\pi $, the lowest
boundary $\varepsilon _{k}^{L}$ and the upper boundary $\varepsilon _{k}^{U}$
of the low-lying excitations both have the same dispersion as the exact ones
$\varepsilon _{Lk}^{Bethe}$ and $\varepsilon _{Uk}^{Bethe}$ of the Bethe
ansatz. Moreover, there emerges a broad mixed region in the frequency $%
\omega $ and momentum $k$ between $\varepsilon _{k}^{L}$ and $\varepsilon
_{k}^{U}$ of the low-lying excitations of the magnons and nearly deconfined
spinons determined by the Eq.(\ref{8a}).

\section{The square lattice}

For the square lattice, according to the Eqs.(\ref{4a1},\ref{4a2}), the
low-lying excitation spectrums of the spins are determined by the following
equations,%
\begin{equation}
\omega ^{2}=\Delta \left( \omega \right) \left[ 2-\zeta _{k}\right] -\Gamma
\left( \omega \right) \left[ 2-\eta _{k}\right]  \label{10a}
\end{equation}%
\begin{equation}
\omega ^{2}=\Delta \left( \omega \right) \left[ 2+\zeta _{k}\right] -\Gamma
\left( \omega \right) \left[ 2-\eta _{k}\right]  \label{10b}
\end{equation}%
where $\zeta _{k}=\cos k_{x}+\cos k_{y}$, and $\eta _{k}=\cos 2k_{x}+\cos
2k_{y}$. These equations are similar to that ones for a chain, and the last
equation (\ref{10b}) will disappear without using the bipartite sublattice
representation. Thus the physical low-lying excitations of the system are
determined by the Eq.(\ref{10a}) which is independent of the bipartite
sublattice representation.

Like that for a chain, the low-lying excitation spectrums is determined by
the Eq.(\ref{10a}), in which there are two kinds of the low-lying
excitations represented by the factors $2-\eta _{k}$, and $2-\zeta _{k}$,
respectively. Around the momentum, $\mathbf{k}=\left( 0,0\right) $, the
possible region of the low-lying excitations in the frequency $\omega $ axis
is narrow, since both the factors $2-\zeta _{k}$ and $2-\eta _{k}$ go to
zero.

In the region around the momentum, $\mathbf{k}=\left( \pi ,\pi \right) $,
the low-lying excitations have a broad distribution in the frequency $\omega
$ axis, and their upper and the lowest boundaries are determined by the Eq.(%
\ref{10a}). In the high energy limit, $\omega /J\rightarrow \infty $, the
coefficient $\Delta \left( \omega \right) $ is a constant, $\Delta \left(
\omega \right) =J^{2}$, while the coefficient $\Gamma \left( \omega \right) $
goes to zero, $\Gamma \left( \omega \right) =0$, thus we obtain the upper
boundary of the low-lying excitations,%
\begin{equation}
\varepsilon ^{U}(k)=J\left[ 2-\zeta _{k}\right] ^{1/2}  \label{11a}
\end{equation}%
which takes the maximum value at $\mathbf{k}=\left( \pi ,\pi \right) $, and
it describes the low-lying excitations of nearly deconfined spinons, like
that for 1D. In the low energy limit, $\omega /J\rightarrow 0$, the
coefficient $\Delta \left( \omega \right) $ goes to zero, $\Delta \left(
0\right) =0$, and the coefficient $\Gamma \left( \omega \right) $ becomes a
constant, $\Gamma \left( 0\right) =-J^{2}/16$. Thus we obtain the lowest
boundary of the low-lying excitations,%
\begin{equation}
\varepsilon ^{L}(k)=\frac{J}{4}\left[ 2-\eta _{k}\right] ^{1/2}  \label{11b}
\end{equation}%
that represents the spin wave excitation of magnons (paired spinons). In the
broad mixed region between $\varepsilon ^{L}(k)$ and $\varepsilon ^{U}(k)$,
there are two kinds of modes of the low-lying excitations, that are
represented by the factors, $2-\zeta _{k}$ and $2-\eta _{k}$, respectively.
In practice, the coefficients $\varepsilon ^{L}(k)$ and $\varepsilon ^{U}(k)$
are modified due to both the coefficients $\Delta \left( \omega \right) $
and $\Gamma \left( \omega \right) $ are the function of the frequency $%
\omega $.

According to the Eq.(\ref{10a}), the lowest boundary of the low-lying
excitations around $\mathbf{k}=\left( \pi ,0\right) $ or $\mathbf{k}=\left(
0,\pi \right) $ is generally different from that ones around $\mathbf{k}%
=\left( \frac{\pi }{2},\frac{\pi }{2}\right) $, even though they have the
same upper boundary of the low-lying excitations represented by $\varepsilon
^{U}(k)$. In the region around $\mathbf{k}=\left( \pi ,0\right) $ or $%
\mathbf{k}=\left( 0,\pi \right) $, these two modes of the low-lying
excitations have slowly varying momentum dependence of the forms $\cos
\Delta k_{x}\pm \cos \Delta k_{y}$ and $\cos 2\Delta k_{x}\pm \cos 2\Delta
k_{y}$, where $\Delta k_{x}$ and $\Delta k_{y}$ are small quantities away
from the point $\mathbf{k}=\left( \pi ,0\right) $ or $\mathbf{k}=\left(
0,\pi \right) $. While in the region around $\mathbf{k}=\left( \frac{\pi }{2}%
,\frac{\pi }{2}\right) $, these low-lying excitations have the momentum
dependence of the forms $\sin \Delta k_{x}+\sin \Delta k_{y}$ and $\cos
2\Delta k_{x}+\cos 2\Delta k_{y}$. The low-lying excitations in these two
different regions have a distinct symmetry about $\Delta k_{x}$ and $\Delta
k_{y}$. On the other hand, in these two different regions, the difference
between the upper and the lowest boundaries of the low-lying excitation is
much less than that in the region around $\mathbf{k}=\left( \pi ,\pi \right)
$. These prominent characters of the low-lying excitations in the short
wave-length regions of the Brillouin zone have been observed in recent
neutron scattering observations\cite{12} and numerical calculations\cite{13}
where this phenomenon is explained as nearly deconfined multispinon
excitations.

The EOMs of spin susceptibility in the Eqs.(\ref{4a1},\ref{4a2}) for a spin $%
1/2$ Heisenberg model are independent of the dimensions of the system, and
they valid for a chain and a square lattice. According to the explanation of
the elementary excitations for a 1D spin $1/2$ Heisenberg model\cite{9}, it
is convincible to believe that the lowest boundary of the low-lying
excitations corresponds to the spin wave excitation of magnons (paired
spinons), and the upper boundary of the low-lying excitations describes the
excitations of nearly deconfied spinons, while in the broad mixed region
between the lowest and upper boundaries the low-lying excitations there
exists a cross-over from magnons (paired spinons) close to the lowest
boundary to nearly deconfined spinons near the upper boundary. It is
reasonable to conjecture that the coupling strength between two spinons is
decreased as the frequency $\omega $ increasing from the lowest boundary to
upper boundary of the low-lying excitations. This explanation of the
low-lying excitations of the system is consistent with the experimental
observations\cite{12} and numerical calculations\cite{13}.

\section{The honeycomb lattice}

For the honeycomb lattice, there naturally exists the bipartite sublattice
structure, in the high energy limit, $\omega /J\rightarrow \infty $, the
coefficients in Eqs.(\ref{4a1},\ref{4a2}) take the values, $\Delta \left(
\omega \right) =J^{2}$, and $\Gamma _{lij}\left( \omega \right) =0$,
respectively, thus we obtain the upper boundary of the low-lying excitation,%
\begin{equation}
E^{U}(k)=\frac{\sqrt{2}J}{2}\left\{ 3+\sqrt{3+\xi _{k}}\right\} ^{1/2}
\label{12a}
\end{equation}%
where $\xi _{k}=2\cos \sqrt{3}k_{y}+4\cos \frac{3k_{x}}{2}\cos \frac{\sqrt{3}%
k_{y}}{2}$. The low-lying excitation spectrum $E^{U}(k)$ takes the maximum
values at the points $\mathbf{k}=\left( 0,0\right) $ and $\mathbf{k}=\frac{%
2\pi }{3}\left( \pm 1,\pm \sqrt{3}\right) $, respectively.

In the low energy limit, $\omega /J\rightarrow 0$, according to the Eqs.(\ref%
{4a1},\ref{4a2}), we obtain the following equation that can be used to
determine the lowest boundary of the low-lying excitations,
\begin{equation}
\omega ^{2}=-\frac{J^{2}}{4}\left[ 3-\sqrt{3+\xi _{k}}\right] +\frac{J^{2}}{%
16}\left[ 6-\xi _{k}\right]  \label{12b}
\end{equation}%
Obviously, this equation has real solutions only in the very small regions
around the high symmetry points of the honeycomb lattice, such as, $\mathbf{k%
}=\left( 0,0\right) $, $\mathbf{k}=\frac{2\pi }{3}\left( \pm 1,\pm \sqrt{3}%
\right) $, because the coefficient of the first term in the right hand side
is negative. On the other hand, it has not real solutions in the regions
around the Dirac points $\mathbf{k}=\frac{2\pi }{3}\left( \pm 1,\pm \frac{%
\sqrt{3}}{3}\right) $, thus the low-lying excitations are gapful these Dirac
points. However, in the mixed region between the upper and the lowest
boundaries of the low-lying excitations, there is only one mode of the
low-lying excitations represented by the factor $\xi _{k}$, which is
different from that ones for a chain and a square lattice. These prominent
characters for the honeycomb lattice can be tested in the future neutron
scattering observations and numerical calculations.

\section{Conclusion and discussion}

For a spin chain, we have shown that the upper and the lowest boundaries of
the low-lying excitations have the same dispersion in the whole range of
momentum as that exact ones obtained by the Bethe ansatz for the $S=1$
excitation of the spinon pairs. For a square lattice, there are two modes of
the low-lying excitations, and they coexist in a broad region between the
upper and the lowest boundaries of the low-lying excitation of spins in the
BZ. The lowest boundary of excitations corresponds the usual spin wave
which mainly takes place in the region around $\mathbf{k}=\left( \pi ,\pi
\right) $, while in other regions, such as around $\mathbf{k}=\left( \pi
,0\right) $ and $\mathbf{k}=\left( \frac{\pi }{2},\frac{\pi }{2}\right) $,
it is very weak. Usually it is seen as the low-lying excitations of magnons
(paired spinons). Another one mode resides mainly in the higher energy
and/or short wave-length regions, such as around $\mathbf{k}=\left( \pi
,0\right) $ and $\mathbf{k}=\left( \frac{\pi }{2},\frac{\pi }{2}\right) $.
It is called the low-lying excitations of nearly deconfined spinons. In the
mixed region between the lowest and upper boundaries of the low-lying
excitations there is a cross-over of the elementary excitations from magnons
close to the lowest boundary to nearly deconfined spinons near the upper
boundary. For a honeycomb lattice, the upper and the lowest boundaries of
the low lying excitations can be described by one mode of the low-lying
excitation which is represented by the factor $\xi _{k}$, where the upper
boundary of spectrum is gapful in whole BZ, and the lowest boundary of
spectrum has the zero points at the high symmetry points $\mathbf{k}=\left(
0,0\right) $ and $\mathbf{k}=\frac{2\pi }{3}\left( \pm 1,\pm \sqrt{3}\right)
$ of the honeycomb lattice.

All of these prominent characters of the low-lying excitation are
independent of whether the system has a long range order that may modify the
spectral weight of the low-lying excitations of the system, and they are
completely determined by the local SU(2) symmetry of the spin $1/2$
antiferromagnetic Heisenberg model and the structure of the lattice of the
spins residing in, such as, a square lattice, a honeycomb lattice, or
others. However, the parameter $<\widehat{s}_{i}^{z}>$ appearing in the
equation of motion of the spin susceptibility can be used to judge whether
the system has a long range order. The recent neutron scattering observations%
\cite{12} have clearly shown that in the regime around $\mathbf{k}=\left(
\pi ,\pi \right) $ the spectral function of the low-lying excitations has
anomalously broad peaks, and this exotic behaviour of the low-lying
excitations can be reasonably explained by the present calculations.
Moreover, the present results obey the Lieb-Schultz-Mattis theorem and its
generalizations, and for a chain, the upper and the lowest boundaries of the
low-lying excitation have the same dispersion as that ones of the Bethe
ansatz, only their coefficients are different, which can be modified by
including the contributions of the high order multiple-point correlation
functions.

Just as shown\cite{9} for a chain, the elementary low-lying excitations are
spinons, and they carry fractional spin ($S=1/2$) which restricts them to
being created in (multiple) pairs. The lowest boundary of the low-lying
excitations is the spectrum of the two spinon excitation with the spin $S=1$
(magnon). However, the upper boundary of the low-lying excitation has a
different dispersion from that of this excitation spectrum of the two spinon
excitation, where its period is twice that of the latter one. It is well
known that the upper boundary of the low-lying excitations is the spectrum
of nearly deconfined multispinon (pairs) excitations\cite%
{12a,12b,12c,12d,12e}. For a square lattice, it has the similar excitation
spectrums like that for a chain, where the lowest boundary of the low-lying
excitation around $\mathbf{k}=\left( \pi ,\pi \right) $ represents the spin
wave excitation represented by the factor $2-\eta _{k}$, while as away from
this region, the spin wave excitation is strongly suppressed, and it becomes
very weak, since the coefficient of the factor $2-\eta _{k}$ rapidly
decreasing as the frequency $\omega $ increasing. In a broad higher energy
region between the upper and the lowest boundaries around the momentum $%
\mathbf{k}=\left( \pi ,\pi \right) $ and other short wave-length regions,
where the spin wave excitation nearly disappears, there emerges another
low-lying excitation represented by the factor $2-\zeta _{k}$, where its
coefficient rapidly increasing as the frequency $\omega $ increasing. that
mainly contributes to the spectral weight of the low-lying excitations in
these regions. These our results are completely consistent with the neutron
scattering observations and numerical calculations for the square lattice.
For the honeycomb lattice, the upper and the lowest boundaries of the low
lying excitations of the spins can be described by one mode represented by
the factor $\xi _{k}$, that can be tested in the future neutron scattering
experiments.

\section{Acknowledgments}

This work is supported by the National Natural Science Foundation of China
under Grant No. 11074301 and No. 11974420, and the National Basic Research
Program of China under Grant No. 2012CB921704.

\subsection{Appendix A: Equations of motion of the spin susceptibility and
related multiple-point correlation functions}

The spin $S=1/2$ antiferromagnetic Heisenberg model is defined by the
Hamiltonian on the bipartite sublattice,%
\begin{equation}
H=\sum_{<i,j>}\left[ J_{ij}^{\bot }\left( \widehat{s}_{i}^{x}\widehat{\tau }%
_{j}^{x}+\widehat{s}_{i}^{y}\widehat{\tau }_{j}^{y}\right) +J_{ij}^{z}%
\widehat{s}_{i}^{z}\widehat{\tau }_{j}^{z}\right]   \label{1}
\end{equation}%
where the antiferromagnetic exchanges $J_{ij}^{\bot }$ (here we take $%
J_{ij}^{x}=J_{ij}^{y}=J_{ij}^{\bot }$) and $J_{ij}^{z}$ are restricted to
nearest neighbor spins $<i,j>$, $J_{ij}^{\bot }=J^{\bot }$, $J_{ij}^{z}=J^{z}
$, other cases, they are zero, and for the isotropic case, $J^{\bot }=J^{z}=J
$. The spin operators $\widehat{\mathbf{s}}_{i}$ and $\widehat{\mathbf{\tau }%
}_{i}$ satisfy the commutation relations ($\hbar =1$),%
\begin{equation}
\left[ \widehat{o}_{i}^{\mu },\widehat{o}_{j}^{\nu }\right] =i\delta
_{ij}\epsilon _{\mu \nu \lambda }\widehat{o}_{i}^{\lambda }  \label{2}
\end{equation}%
where $\delta _{ij}$ is the Kronecker delta function, and $\epsilon _{\mu
\nu \lambda }$ is an antisymmetry tensor, $\epsilon _{xyz}=1$. With the
Heisenberg representation, the time dependence of the spin operators is that,%
\begin{equation}
\widehat{o}_{i}^{\mu }(t)=e^{iHt}\widehat{o}_{i}^{\mu }e^{-iHt}  \label{3}
\end{equation}%
where the operator $\widehat{\mathbf{o}}_{i}=\widehat{\mathbf{s}}_{i},%
\widehat{\mathbf{\tau }}_{i}$.

According to the algebraic equation of motion approach, we need the
following commutation relations,

\begin{eqnarray}
\left[ \widehat{s}_{i}^{z},H\right]  &=&\frac{1}{2}J_{im}^{\bot }\widehat{X}%
_{im}^{\left( -\right) }  \notag \\
\left[ \widehat{\tau }_{i}^{z},H\right]  &=&-\frac{1}{2}J_{im}^{\bot }%
\widehat{X}_{mi}^{\left( -\right) }  \label{aa2a}
\end{eqnarray}%
\begin{eqnarray}
\left[ \widehat{s}_{i}^{-},H\right]  &=&-J_{im}^{\perp }\widehat{\tau }%
_{m}^{-}\widehat{s}_{i}^{z}+J_{im}^{z}\widehat{\tau }_{m}^{z}\widehat{s}%
_{i}^{-}  \notag \\
\left[ \widehat{\tau }_{i}^{-},H\right]  &=&-J_{im}^{\perp }\widehat{s}%
_{m}^{-}\widehat{\tau }_{i}^{z}+J_{im}^{z}\widehat{s}_{m}^{z}\widehat{\tau }%
_{i}^{-}  \label{aa2b}
\end{eqnarray}%
\begin{eqnarray}
\left[ \widehat{X}_{ij}^{\left( \mp \right) },H\right]  &=&J_{im}^{\perp }%
\widehat{\Lambda }_{mj}^{\left( \pm \right) }\widehat{s}_{i}^{z}-J_{jm}^{%
\perp }\widehat{\Gamma }_{im}^{\left( \pm \right) }\widehat{\tau }_{j}^{z}
\notag \\
&&-J_{im}^{z}\widehat{X}_{ij}^{\left( \pm \right) }\widehat{\tau }%
_{m}^{z}+J_{jm}^{z}\widehat{X}_{ij}^{\left( \pm \right) }\widehat{s}%
_{m}^{z}+J_{ij}^{z}\widehat{X}_{ij}^{\left( \mp \right) }  \label{aa2c}
\end{eqnarray}%
\begin{equation}
\left[ \widehat{\Gamma }_{ij}^{\left( \pm \right) },H\right] =\mp
J_{im}^{\perp }\widehat{X}_{jm}^{\left( \mp \right) }\widehat{s}%
_{i}^{z}-J_{jm}^{\perp }\widehat{X}_{im}^{\left( \mp \right) }\widehat{s}%
_{j}^{z}-\left( J_{im}^{z}-J_{jm}^{z}\right) \widehat{\Gamma }_{ij}^{\left(
\mp \right) }\widehat{\tau }_{m}^{z}  \label{aa2d}
\end{equation}%
\begin{equation}
\left[ \widehat{\Lambda }_{ij}^{\left( \pm \right) },H\right] =J_{im}^{\perp
}\widehat{X}_{mj}^{\left( \mp \right) }\widehat{\tau }_{i}^{z}\pm
J_{jm}^{\perp }\widehat{X}_{mi}^{\left( \mp \right) }\widehat{\tau }%
_{j}^{z}-\left( J_{im}^{z}-J_{jm}^{z}\right) \widehat{\Lambda }_{ij}^{\left(
\mp \right) }\widehat{s}_{m}^{z}  \label{aa2e}
\end{equation}%
where $\widehat{X}_{ij}^{\left( \pm \right) }=\widehat{s}_{i}^{+}\widehat{%
\tau }_{j}^{-}\pm \widehat{\tau }_{j}^{+}\widehat{s}_{i}^{-}$, $\widehat{%
\Gamma }_{ij}^{\left( \pm \right) }=\widehat{s}_{i}^{+}\widehat{s}%
_{j}^{-}\pm \widehat{s}_{j}^{+}\widehat{s}_{i}^{-}$, and $\widehat{\Lambda }%
_{ij}^{\left( \pm \right) }=\widehat{\tau }_{i}^{+}\widehat{\tau }%
_{j}^{-}\pm \widehat{\tau }_{j}^{+}\widehat{\tau }_{i}^{-}$. These
commutation relations are the basic ingredients to writing equations of
motion of high order correlation functions.

The transverse and longitudinal spin susceptibilities are defined as that,%
\begin{equation}
\chi _{ij}^{\mu \nu }(t)=i\theta (t)<\left[ \widehat{s}_{i}^{\mu }(t),%
\widehat{s}_{j}^{\nu }(0)\right] >  \label{aa3a}
\end{equation}%
\begin{equation}
\widetilde{\chi }_{ij}^{\mu \nu }(t)=i\theta (t)<\left[ \widehat{\tau }%
_{i}^{\mu }(t),\widehat{s}_{j}^{\nu }(0)\right] >  \label{aa3b}
\end{equation}%
where $\mu ,\nu =\pm ,z$. In order to tersely represent the equation of
motion of the spin susceptibility, we define the following correlation
functions that some of them appearing in the hierarchic series of EOM of the
spin susceptibility, called related multiple-point correlation functions,%
\begin{eqnarray}
\widetilde{F}_{\left\{ \alpha \right\} jq}^{(\left\{ A\right\} )}(t)
&=&i\theta (t)<\left[ \Pi _{k=1}^{N}\left[ \widehat{A}_{\alpha }(t)\right]
^{k}\widehat{\tau }_{j}^{-}(t),\widehat{s}_{q}^{+}(0)\right] >  \notag \\
F_{\left\{ \alpha \right\} jq}^{(\left\{ A\right\} )}(t) &=&i\theta (t)<
\left[ \Pi _{k=1}^{N}\left[ \widehat{A}_{\alpha }(t)\right] ^{k}\widehat{s}%
_{j}^{-}(t),\widehat{s}_{q}^{+}(0)\right] >  \label{aa4a}
\end{eqnarray}%
\begin{eqnarray}
\widetilde{L}_{\left\{ \alpha \right\} jq}^{(\left\{ A\right\} )}(t)
&=&i\theta (t)<\left[ \Pi _{k=1}^{N}\left[ \widehat{A}_{\alpha }(t)\right]
^{k}\widehat{\tau }_{j}^{z}(t),\widehat{s}_{q}^{z}(0)\right] >  \notag \\
L_{\left\{ \alpha \right\} jq}^{(\left\{ A\right\} )}(t) &=&i\theta (t)<
\left[ \Pi _{k=1}^{N}\left[ \widehat{A}_{\alpha }(t)\right] ^{k}\widehat{s}%
_{j}^{z}(t),\widehat{s}_{q}^{z}(0)\right] >  \label{aa4b}
\end{eqnarray}%
where $\widehat{A}_{\alpha }=\widehat{s}_{i}^{z},\widehat{\tau }_{i}^{z},%
\widehat{X}_{ij}^{\left( \pm \right) },\widehat{\Gamma }_{ij}^{\left( \pm
\right) },\widehat{\Lambda }_{ij}^{\left( \pm \right) }$, and the parameter $%
N$ represents the level of the corresponding correlation function in the
hierarchic structure of the EOM of the spin susceptibility. For the
longitudinal spin susceptibility $\chi _{ij}^{zz}(t)$ and $\widetilde{\chi }%
_{ij}^{zz}(t)$, we need to define another special related multiple-point
correlation function, $K_{ijq}(t)=i\theta (t)<\left[ \widehat{X}%
_{ij}^{\left( -\right) }(t),\widehat{s}_{q}^{z}(0)\right] >$, that belonging
to the $N=1$ level.

With the commutation relations in Eqs(\ref{aa2a},\ref{aa2b}), we can write
out the following the EOMs of the spin susceptibility after taking the
Fourier transformation of time,

\begin{eqnarray}
\omega \chi _{iq}^{-+}(\omega ) &=&2s^{z}\delta _{iq}-\sum_{m}J_{im}^{\perp }%
\widetilde{F}_{imq}^{(s)}(\omega )+\sum_{m}J_{im}^{z}F_{miq}^{(\tau
)}(\omega )  \notag \\
\omega \widetilde{\chi }_{iq}^{-+}(\omega ) &=&-\sum_{m}J_{im}^{\perp
}F_{imq}^{(\tau )}(\omega )+\sum_{m}J_{im}^{z}\widetilde{F}%
_{miq}^{(s)}(\omega )  \label{5a}
\end{eqnarray}%
\begin{eqnarray}
\omega \chi _{iq}^{zz}(\omega ) &=&\frac{1}{2}\sum_{m}J_{im}^{\bot
}K_{imq}(\omega )  \notag \\
\omega \widetilde{\chi }_{iq}^{zz}(\omega ) &=&-\frac{1}{2}%
\sum_{m}J_{im}^{\bot }K_{miq}(\omega )  \label{5b}
\end{eqnarray}%
where $s^{z}=<\widehat{s}_{i}^{z}>$. The related multiple-point correlation
functions $\widetilde{F}_{ijq}^{(s)}(\omega )$, $F_{ijq}^{(\tau )}(\omega )$
and $K_{ijq}(\omega )$ belong to the $N=1$ level, and with the help of the
relations of the spin operators, $\left( \widehat{s}_{i}^{z}\right)
^{2}=\left( \widehat{\tau }_{i}^{z}\right) ^{2}=\frac{1}{4}$, $\widehat{s}%
_{i}^{+}\widehat{s}_{i}^{-}=\frac{1}{2}+\widehat{s}_{i}^{z}$ and $\widehat{%
\tau }_{i}^{+}\widehat{\tau }_{i}^{-}=\frac{1}{2}+\widehat{\tau }_{i}^{z}$,
their EOMs can be significantly simplified as that,%
\begin{eqnarray}
\omega \widetilde{F}_{ijq}^{(s)}(\omega ) &=&-<\widehat{\tau }_{j}^{-}%
\widehat{s}_{i}^{+}>\delta _{iq}-\frac{1}{4}J_{ij}^{\bot }\chi
_{iq}^{-+}(\omega )+\frac{1}{4}J_{ij}^{z}\widetilde{\chi }_{jq}^{-+}(\omega )
\notag \\
&&-\sum_{m}J_{jm}^{\bot }\left( 1-\delta _{mi}\right) F_{ijmq}^{(s\tau
)}(\omega )+\sum_{m}J_{jm}^{z}\left( 1-\delta _{mi}\right) \widetilde{F}%
_{imjq}^{(ss)}(\omega )  \label{6a} \\
&&+\frac{1}{2}\sum_{m}J_{im}^{\bot }\left( 1-\delta _{mj}\right) \widetilde{F%
}_{imjq}^{(X^{\left( -\right) })}(\omega )  \notag
\end{eqnarray}%
\begin{eqnarray}
\omega F_{ijq}^{(\tau )}(\omega ) &=&2<\widehat{\tau }_{i}^{z}\widehat{s}%
_{j}^{z}>\delta _{jq}-\frac{1}{4}J_{ij}^{\bot }\widetilde{\chi }%
_{iq}^{-+}(\omega )+\frac{1}{4}J_{ij}^{z}\chi _{jq}^{-+}(\omega )  \notag \\
&&-\sum_{m}J_{jm}^{\bot }\left( 1-\delta _{mi}\right) \widetilde{F}%
_{ijmq}^{(\tau s)}(\omega )+\sum_{m}J_{jm}^{z}\left( 1-\delta _{mi}\right)
F_{imjq}^{(\tau \tau )}(\omega )  \label{6b} \\
&&-\frac{1}{2}\sum_{m}J_{im}^{\bot }\left( 1-\delta _{mj}\right)
F_{mijq}^{(X^{\left( -\right) })}(\omega )  \notag
\end{eqnarray}%
\begin{eqnarray}
\omega K_{ijq}(\omega ) &=&<\widehat{X}_{ij}^{\left( +\right) }>\delta
_{iq}+J_{ij}^{\perp }\chi _{iq}^{zz}(\omega )-J_{ij}^{\perp }\widetilde{\chi
}_{jq}^{zz}(\omega )  \notag \\
&&+\sum_{m}J_{im}^{\perp }\left( 1-\delta _{mj}\right) L_{mjiq}^{\left(
\Lambda ^{+}\right) }(\omega )-\sum_{m}J_{jm}^{\perp }\left( 1-\delta
_{mi}\right) \widetilde{L}_{imjq}^{\left( \Gamma ^{+}\right) }(\omega )
\label{6c} \\
&&-\sum_{m}J_{im}^{z}\left( 1-\delta _{mj}\right) \widetilde{L}%
_{ijmq}^{\left( X^{+}\right) }(\omega )+\sum_{m}J_{jm}^{z}\left( 1-\delta
_{mi}\right) L_{ijmq}^{\left( X^{+}\right) }(\omega )  \notag
\end{eqnarray}%
Notice that the spin susceptibilities $\chi _{iq}^{-+}(\omega )$ and $%
\widetilde{\chi }_{jq}^{-+}(\omega )$ appear in the above equations without
taking any approximation, which is a key character of the algebraic equation
of motion approach to the spin $S=1/2$ magnetic systems. The related
multiple-point correlation functions, such as $\widetilde{F}%
_{imjq}^{(X^{\left( -\right) })}(\omega )$, $\widetilde{F}_{ijmq}^{(\tau
s)}(\omega )$, $F_{imjq}^{(\tau \tau )}(\omega )$, and $L_{mjiq}^{\left(
\Lambda ^{+}\right) }(\omega )$, \textit{et al}., belong to the $N=2$ level
in the hierarchic series of the EOMs of the spin susceptibility.

\subsection{Appendix B: The XY model}

For the XY model, to studying its low-lying excitations, we need to solve
the EOMs of the longitudinal spin susceptibilities $\chi _{ij}^{zz}(t)$ and $%
\widetilde{\chi }_{ij}^{zz}(t)$, in which there only appears the
multiple-point correlation function $K_{ijq}(\omega )$ that is determined by
the Eq.(\ref{6c}) with $J_{ij}^{z}=0$. Using the expression of the $%
K_{ijq}(\omega )$, we can rewrite out the EOMs of the longitudinal spin
susceptibilities $\chi _{ij}^{zz}(t)$ and $\widetilde{\chi }_{ij}^{zz}(t)$
as that,
\begin{eqnarray}
\left[ \omega ^{2}-\eta \right] \chi _{iq}^{zz}(\omega )
&=&\sum_{m}J_{im}^{\bot }<\widehat{X}_{im}^{\left( +\right) }>\delta _{iq}-%
\frac{1}{2}\sum_{m}\left( J_{im}^{\bot }\right) ^{2}\widetilde{\chi }%
_{mq}^{zz}(\omega )  \notag \\
&&-\frac{1}{2}\sum_{mn}J_{im}^{\bot }J_{mn}^{\perp }\left( 1-\delta
_{ni}\right) \widetilde{L}_{inmq}^{\left( \Gamma ^{+}\right) }(\omega )+%
\frac{1}{2}\sum_{mn}J_{im}^{\bot }J_{in}^{\perp }\left( 1-\delta
_{nm}\right) L_{nmiq}^{\left( \Lambda ^{+}\right) }(\omega )  \label{ba}
\end{eqnarray}%
\begin{eqnarray}
\left[ \omega ^{2}-\eta \right] \widetilde{\chi }_{iq}^{zz}(\omega )
&=&-\sum_{m}J_{im}^{\bot }<\widehat{X}_{mi}^{\left( +\right) }>\delta _{mq}-%
\frac{1}{2}\sum_{m}\left( J_{im}^{\bot }\right) ^{2}\chi _{mq}^{zz}(\omega )
\notag \\
&&-\frac{1}{2}\sum_{mn}J_{im}^{\bot }J_{mn}^{\perp }\left( 1-\delta
_{ni}\right) L_{nimq}^{\left( \Lambda ^{+}\right) }(\omega )+\frac{1}{2}%
\sum_{mn}J_{im}^{\bot }J_{in}^{\perp }\left( 1-\delta _{nm}\right)
\widetilde{L}_{mniq}^{\left( \Gamma ^{+}\right) }(\omega )  \label{bb}
\end{eqnarray}%
where $\eta =\frac{1}{2}\sum_{m}\left( J_{im}^{\bot }\right) ^{2}$. As a
simple approximation, we discard the $L_{nmiq}^{\left( \Lambda ^{+}\right)
}(\omega )$ term in the Eq.(\ref{ba}), and the $\widetilde{L}_{mniq}^{\left(
\Gamma ^{+}\right) }(\omega )$ term in the Eq.(\ref{bb}), respectively. The
reason is that, for example, according to the definition of the correlation
function $L_{nmiq}^{\left( \Lambda ^{+}\right) }(\omega )$, it represents a
time evolution of a spin operator $\widehat{s}_{i}^{z}(t)$ with its neighbor
$\widehat{\Lambda }_{nm}^{\left( +\right) }(t)$ from an initial state at
time $t$ to a final state at time $t^{\prime }=0$. Since the $%
L_{nmiq}^{\left( \Lambda ^{+}\right) }(\omega )$ has the same label $i$ as
that of the spin susceptibility $\chi _{iq}^{zz}(\omega )$, the $%
L_{nmiq}^{\left( \Lambda ^{+}\right) }(\omega )$ term only describes the
influence of other spins around the spin operator $\widehat{s}_{i}^{z}(t)$
on the spin susceptibility $\chi _{iq}^{zz}(\omega )$, and it does not
directly represent a spin flipping process of the spin operator $\widehat{%
\mathbf{s}}_{i}(t)$. As compared with the $\widetilde{L}_{inmq}^{\left(
\Gamma ^{+}\right) }(\omega )$ term, its contribution to the spin
susceptibility $\chi _{iq}^{zz}(\omega )$ can be neglected. Under these
approximations, the above equations are rewritten as that,%
\begin{eqnarray}
\left[ \omega ^{2}-\eta \right] \chi _{iq}^{zz}(\omega )
&=&\sum_{m}J_{im}^{\bot }<\widehat{X}_{im}^{\left( +\right) }>\delta _{iq}-%
\frac{1}{2}\sum_{m}\left( J_{im}^{\bot }\right) ^{2}\widetilde{\chi }%
_{mq}^{zz}(\omega )  \notag \\
&&-\frac{1}{2}\sum_{mn}J_{im}^{\bot }J_{mn}^{\perp }\left( 1-\delta
_{ni}\right) \widetilde{L}_{inmq}^{\left( \Gamma ^{+}\right) }(\omega )
\label{ba1}
\end{eqnarray}%
\begin{eqnarray}
\left[ \omega ^{2}-\eta \right] \widetilde{\chi }_{iq}^{zz}(\omega )
&=&-\sum_{m}J_{im}^{\bot }<\widehat{X}_{mi}^{\left( +\right) }>\delta _{mq}-%
\frac{1}{2}\sum_{m}\left( J_{im}^{\bot }\right) ^{2}\chi _{mq}^{zz}(\omega )
\notag \\
&&-\frac{1}{2}\sum_{mn}J_{im}^{\bot }J_{mn}^{\perp }\left( 1-\delta
_{ni}\right) L_{nimq}^{\left( \Lambda ^{+}\right) }(\omega )  \label{bb1}
\end{eqnarray}%
that are used to calculated the spin susceptibility of the spin $1/2$ XY
model.

With the help of the Eqs.(\ref{aa2d},\ref{aa2e}), we can write out the EOMs
of the multiple-point correlation functions $L_{ijlq}^{\left( \Lambda
^{+}\right) }(\omega )$ and $\widetilde{L}_{ijlq}^{\left( \Gamma ^{+}\right)
}(\omega )$ ($J_{ij}^{z}=0$),%
\begin{equation}
\omega L_{mjiq}^{\left( \Lambda ^{+}\right) }(\omega )=\sum_{n}\left[ \frac{1%
}{2}J_{in}^{\bot }K_{mjinq}^{\left( \Lambda ^{+}X^{-}\right) }(\omega
)+J_{mn}^{\perp }L_{njmiq}^{\left( X^{-}\tau \right) }(\omega
)+J_{jn}^{\perp }L_{nmjiq}^{\left( X^{-}\tau \right) }(\omega )\right]
\label{b1a}
\end{equation}%
\begin{equation}
\omega \widetilde{L}_{imjq}^{\left( \Gamma ^{+}\right) }(\omega )=-\sum_{n}%
\left[ \frac{1}{2}J_{jn}^{\bot }K_{imnjq}^{\left( \Gamma ^{+}X^{-}\right)
}(\omega )+J_{in}^{\perp }\widetilde{L}_{mnijq}^{\left( X^{-}Z\right)
}(\omega )+J_{mn}^{\perp }\widetilde{L}_{inmjq}^{\left( X^{-}Z\right)
}(\omega )\right]  \label{b1b}
\end{equation}%
where we have neglected the static quantities appearing in these EOMs.

As writing out the summation over the lattice sites in the right side of the
above EOMs, there may appear some multiple-point correlation functions
belonging to the $N=3$ level that have two same labels, such as, $%
K_{mjimq}^{\left( \Lambda ^{+}X^{-}\right) }(\omega )$, $K_{imijq}^{\left(
\Gamma ^{+}X^{-}\right) }(\omega )$, $L_{ijmiq}^{\left( X^{-}\tau \right)
}(\omega )$, \textit{et al}.. Using the relations $\left( \widehat{s}%
_{i}^{z}\right) ^{2}=\left( \widehat{\tau }_{i}^{z}\right) ^{2}=\frac{1}{4}$%
, $\widehat{s}_{i}^{+}\widehat{s}_{i}^{-}=\frac{1}{2}+\widehat{s}_{i}^{z}$
and $\widehat{\tau }_{i}^{+}\widehat{\tau }_{i}^{-}=\frac{1}{2}+\widehat{%
\tau }_{i}^{z}$, to simplify these multiple-point correlation functions
where there emerge some ones belonging to the $N=1$ level, and finally
discarding multiple-point correlation functions belonging to the $N=3$ level%
\cite{19} (called a "soft cut-off" approximation), we can rewrite out the
Eqs.(\ref{b1a},\ref{b1b}) as that,%
\begin{equation}
\omega \widetilde{L}_{imjq}^{\left( \Gamma ^{+}\right) }(\omega )=-\frac{1}{4%
}J_{jm}^{\bot }K_{ijq}(\omega )-\frac{1}{4}J_{ji}^{\bot }K_{mjq}(\omega )
\label{b2a}
\end{equation}%
\begin{equation}
\omega L_{mjiq}^{\left( \Lambda ^{+}\right) }(\omega )=\frac{1}{4}%
J_{im}^{\bot }K_{ijq}(\omega )+\frac{1}{4}J_{ij}^{\bot }K_{imq}(\omega )
\label{b2b}
\end{equation}%
Now the set of equations composed of the Eqs.(\ref{b2a},\ref{b2b}) and Eq.(%
\ref{6c}) are closed, while they are still difficult to be solved, since it
is in fact a set of tensor equations.

Here we approximately solve these equations: (a) Substituting the Eq.(\ref%
{6c}) into the Eq.(\ref{b2a}), we discard the $L_{mjiq}^{\left( \Lambda
^{+}\right) }(\omega )$ term  or substituting the Eq.(\ref{6c}) into the Eq.(%
\ref{b2b}), we discard the $\widetilde{L}_{imjq}^{\left( \Gamma ^{+}\right)
}(\omega )$ term; Consequently, we in fact discard the coupling between the
multiple-point correlation functions $\widetilde{L}_{imjq}^{\left( \Gamma
^{+}\right) }(\omega )$ and $L_{mjiq}^{\left( \Lambda ^{+}\right) }(\omega )$%
, and we bring the set of equations composed of the Eqs.(\ref{b2a},\ref{b2b}%
) and Eq.(\ref{6c}) into two subset of equations. (b) In each subset of
equations, we discard the multiple-point correlation functions that have the
different labels with $\widetilde{L}_{imjq}^{\left( \Gamma ^{+}\right)
}(\omega )$ and $L_{mjiq}^{\left( \Lambda ^{+}\right) }(\omega )$,
respectively, then we can obtain the following solutions of $\widetilde{L}%
_{imjq}^{\left( \Gamma ^{+}\right) }(\omega )$ and $L_{mjiq}^{\left( \Lambda
^{+}\right) }(\omega )$,
\begin{equation*}
\widetilde{L}_{imjq}^{\left( \Gamma ^{+}\right) }(\omega )=\frac{%
J_{ij}^{\bot }J_{jm}^{\perp }\left( 1-\delta _{mi}\right) }{2\left( \omega
^{2}-\frac{\left( J^{\bot }\right) ^{2}}{2}\right) }\left\{ \widetilde{\chi }%
_{jq}^{zz}(\omega )-\frac{1}{2}\left[ \chi _{iq}^{zz}(\omega )+\chi
_{mq}^{zz}(\omega )\right] \right\}
\end{equation*}%
\begin{equation}
L_{mjiq}^{\left( \Lambda ^{+}\right) }(\omega )=\frac{J_{ij}^{\bot
}J_{im}^{\perp }\left( 1-\delta _{mj}\right) }{2\left( \omega ^{2}-\frac{%
\left( J^{\bot }\right) ^{2}}{2}\right) }\left\{ \chi _{iq}^{zz}(\omega )-%
\frac{1}{2}\left[ \widetilde{\chi }_{jq}^{zz}(\omega )+\widetilde{\chi }%
_{mq}^{zz}(\omega )\right] \right\}   \label{b3b}
\end{equation}%
where there appear three spin susceptibilities in the above solutions
defined on three neighbor sites.

Substituting the solutions of the multiple-point correlation functions $%
\widetilde{L}_{imjq}^{\left( \Gamma ^{+}\right) }(\omega )$ and $%
L_{mjiq}^{\left( \Lambda ^{+}\right) }(\omega )$ in the Eq.(\ref{b3b}) to
the Eqs.(\ref{ba1},\ref{bb1}), we obtain a set of equations of the spin
susceptibilities $\chi _{iq}^{zz}(\omega )$ and $\widetilde{\chi }%
_{iq}^{zz}(\omega )$. In the low energy limit $\omega /J^{\bot }\rightarrow
0 $, we can obtain the lowest boundary of the low-lying excitations $%
\varepsilon _{k}^{XY}$ in the Eq.(\ref{6}) for the spin $1/2$ XY model as
solving the Eqs.(\ref{ba1},\ref{bb1}).

\subsection{Appendix C: The contribution of the high order related
multiple-point correlation functions}

With the Eqs.(\ref{aa2a}-\ref{aa2c}), we can write out the EOMs of the high
order multiple-point correlation functions appearing in the Eqs.(\ref{6a},%
\ref{6b}) (taking isotropic coupling, $J_{ij}^{\bot }=J_{ij}^{z}=J_{ij}$),%
\begin{eqnarray}
\omega F_{iljq}^{(X^{\left( -\right) })}(\omega ) &=&\sum_{m}\left[
J_{jm}F_{ilmjq}^{(X^{\left( -\right) }\tau )}(\omega )-J_{jm}\widetilde{F}%
_{iljmq}^{(X^{\left( -\right) }s)}(\omega )\right]   \notag \\
&&+\sum_{m}\left[ J_{lm}F_{ilmjq}^{(X^{\left( +\right) }s)}(\omega
)-J_{im}F_{ilmjq}^{(X^{\left( +\right) }\tau )}(\omega
)+J_{il}F_{iljq}^{(X^{\left( -\right) })}(\omega )\right]   \label{7a} \\
&&+\sum_{m}\left[ J_{im}F_{mlijq}^{(\Lambda ^{\left( +\right) }s)}(\omega
)-J_{lm}F_{imljq}^{(\Gamma ^{\left( +\right) }\tau )}(\omega )\right]
\notag
\end{eqnarray}%
\begin{eqnarray}
\omega \widetilde{F}_{iljq}^{(X^{\left( -\right) })}(\omega ) &=&\sum_{m}%
\left[ J_{jm}\widetilde{F}_{ilmjq}^{(X^{\left( -\right) }s)}(\omega
)-J_{jm}F_{iljmq}^{(X^{\left( -\right) }\tau )}(\omega )\right]   \notag \\
&&+\sum_{m}\left[ J_{lm}\widetilde{F}_{ilmjq}^{(X^{\left( +\right)
}s)}(\omega )-J_{im}\widetilde{F}_{ilmjq}^{(X^{\left( +\right) }\tau
)}(\omega )+J_{il}\widetilde{F}_{iljq}^{(X^{\left( -\right) })}(\omega )%
\right]   \label{7b} \\
&&+\sum_{m}\left[ J_{im}\widetilde{F}_{mlijq}^{(\Lambda ^{\left( +\right)
}s)}(\omega )-J_{lm}\widetilde{F}_{imljq}^{(\Gamma ^{\left( +\right) }\tau
)}(\omega )\right]   \notag
\end{eqnarray}%
\begin{eqnarray}
\omega \widetilde{F}_{lijq}^{(ss)}(\omega ) &=&\sum_{m}\left[ J_{jm}%
\widetilde{F}_{ilmjq}^{(sss)}(\omega )-J_{jm}F_{lijmq}^{(ss\tau )}(\omega )%
\right]   \notag \\
&&+\frac{1}{2}\sum_{m}\left[ J_{im}\widetilde{F}_{Limjq}^{(sX^{\left(
-\right) })}(\omega )+J_{lm}\widetilde{F}_{lmijq}^{(X^{\left( -\right)
}s)}(\omega )\right]   \label{7c}
\end{eqnarray}%
\begin{eqnarray}
\omega F_{lijq}^{(\tau \tau )}(\omega ) &=&\sum_{m}\left[ J_{jm}F_{ilmjq}^{(%
\tau \tau \tau )}(\omega )-J_{jm}\widetilde{F}_{lijmq}^{(\tau \tau
s)}(\omega )\right]   \notag \\
&&-\frac{1}{2}\sum_{m}\left[ J_{im}F_{lmijq}^{(\tau X^{\left( -\right)
})}(\omega )+J_{lm}F_{mlijq}^{(\tau X^{\left( -\right) })}(\omega )\right]
\label{7d}
\end{eqnarray}%
\begin{eqnarray}
\omega \widetilde{F}_{lijq}^{(\tau s)}(\omega ) &=&\sum_{m}\left[ J_{jm}%
\widetilde{F}_{Limjq}^{(\tau ss)}(\omega )-J_{jm}F_{lijmq}^{(\tau s\tau
)}(\omega )\right]   \notag \\
&&+\frac{1}{2}\sum_{m}\left[ J_{im}\widetilde{F}_{Limjq}^{(\tau X^{\left(
-\right) })}(\omega )-J_{lm}\widetilde{F}_{mlijq}^{(X^{\left( -\right)
}s)}(\omega )\right]   \label{7e}
\end{eqnarray}%
\begin{eqnarray}
\omega F_{lijq}^{(s\tau )}(\omega ) &=&\sum_{m}\left[ J_{jm}F_{Limjq}^{(s%
\tau \tau )}(\omega )-J_{jm}\widetilde{F}_{lijmq}^{(s\tau s)}(\omega )\right]
\notag \\
&&-\frac{1}{2}\sum_{m}\left[ J_{im}F_{lmijq}^{(sX^{\left( -\right)
})}(\omega )-J_{lm}F_{lmijq}^{(X^{\left( -\right) }\tau )}(\omega )\right]
\label{7f}
\end{eqnarray}%
where we have neglected the static quantities appearing in these EOMs.

Under the "soft cut-off" approximation that applying the relations $\left(
\widehat{s}_{i}^{z}\right) ^{2}=\left( \widehat{\tau }_{i}^{z}\right) ^{2}=%
\frac{1}{4}$, $\widehat{s}_{i}^{+}\widehat{s}_{i}^{-}=\frac{1}{2}+\widehat{s}%
_{i}^{z}$ and $\widehat{\tau }_{i}^{+}\widehat{\tau }_{i}^{-}=\frac{1}{2}+%
\widehat{\tau }_{i}^{z}$, for the correlation functions that having two same
labels appearing in the summations of the right hand side of the Eqs.(\ref%
{7a}-\ref{7f}), and discarding the multiple-point correlation functions
belonging to the $N=3$ level, we can further simplify these equations as
that,%
\begin{eqnarray}
\left[ \omega ^{2}-A_{lij}\right] F_{lijq}^{(X^{\left( -\right) })}(\omega )
&=&\frac{\left( J_{ij}\right) ^{2}}{4}\left( 2-\delta _{jl}\right)
F_{jilq}^{(X^{\left( -\right) })}(\omega )  \notag \\
&&+\frac{J_{ij}J_{il}\left( 1-\delta _{jl}\right) }{8}\left[ 3\widetilde{%
\chi }_{iq}^{-+}(\omega )-2\chi _{jq}^{-+}(\omega )-\chi _{lq}^{-+}(\omega )%
\right]   \label{a4a}
\end{eqnarray}%
\begin{eqnarray}
\left[ \omega ^{2}-A_{lij}\right] \widetilde{F}_{iljq}^{(X^{\left( -\right)
})}(\omega ) &=&\frac{\left( J_{ij}\right) ^{2}}{4}\left( 2-\delta
_{jl}\right) \widetilde{F}_{ijlq}^{(X^{\left( -\right) })}(\omega )  \notag
\\
&&-\frac{J_{ij}J_{il}\left( 1-\delta _{jl}\right) }{8}\left[ 3\chi
_{iq}^{-+}(\omega )-2\widetilde{\chi }_{jq}^{-+}(\omega )-\widetilde{\chi }%
_{lq}^{-+}(\omega )\right]   \label{a4b}
\end{eqnarray}%
\begin{eqnarray}
\left[ \omega ^{2}-B_{lij}\right] \widetilde{F}_{lijq}^{(ss)}(\omega ) &=&-%
\frac{J_{ij}J_{jl}\left( 1-\delta _{il}\right) }{16}\left[ \chi
_{iq}^{-+}(\omega )+\chi _{lq}^{-+}(\omega )-2\widetilde{\chi }%
_{jq}^{-+}(\omega )\right]   \notag \\
&&-\frac{J_{ij}J_{jl}}{8}\left[ F_{ljiq}^{(X^{\left( -\right) })}(\omega
)+F_{ijlq}^{(X^{\left( -\right) })}(\omega )\right]   \label{a4c} \\
&&-\frac{\left( J_{ij}\right) ^{2}}{4}F_{ljiq}^{(s\tau )}(\omega )-\frac{%
\left( J_{jl}\right) ^{2}}{4}F_{ijlq}^{(s\tau )}(\omega )  \notag
\end{eqnarray}%
\begin{eqnarray}
\left[ \omega ^{2}-B_{lij}\right] F_{lijq}^{(\tau \tau )}(\omega ) &=&-\frac{%
J_{ij}J_{jl}\left( 1-\delta _{il}\right) }{16}\left[ \widetilde{\chi }%
_{iq}^{-+}(\omega )+\widetilde{\chi }_{lq}^{-+}(\omega )-2\chi
_{jq}^{-+}(\omega )\right]   \notag \\
&&+\frac{J_{ij}J_{jl}}{8}\left[ \widetilde{F}_{jliq}^{(X^{\left( -\right)
})}(\omega )+\widetilde{F}_{jilq}^{(X^{\left( -\right) })}(\omega )\right]
\label{4d} \\
&&-\frac{\left( J_{ij}\right) ^{2}}{4}\widetilde{F}_{ljiq}^{(\tau s)}(\omega
)-\frac{\left( J_{jl}\right) ^{2}}{4}\widetilde{F}_{ijlq}^{(\tau s)}(\omega )
\notag
\end{eqnarray}%
\begin{eqnarray}
\left[ \omega ^{2}-C_{lij}\right] \widetilde{F}_{lijq}^{(\tau s)}(\omega )
&=&-\frac{J_{ij}J_{il}\left( 1-\delta _{jl}\right) }{16}\left[ \chi
_{iq}^{-+}(\omega )-\widetilde{\chi }_{lq}^{-+}(\omega )\right]   \notag \\
&&-\frac{J_{ij}J_{il}}{8}\widetilde{F}_{iljq}^{(X^{\left( -\right)
})}(\omega )-\frac{\left( J_{ij}\right) ^{2}}{4}F_{ljiq}^{(\tau \tau
)}(\omega )  \label{4e}
\end{eqnarray}%
\begin{eqnarray}
\left[ \omega ^{2}-C_{lij}\right] F_{lijq}^{(s\tau )}(\omega ) &=&-\frac{%
J_{ij}J_{il}\left( 1-\delta _{jl}\right) }{16}\left[ \widetilde{\chi }%
_{iq}^{-+}(\omega )-\chi _{lq}^{-+}(\omega )\right]   \notag \\
&&+\frac{J_{ij}J_{il}}{8}F_{lijq}^{(X^{\left( -\right) })}(\omega )-\frac{%
\left( J_{ij}\right) ^{2}}{4}\widetilde{F}_{ljiq}^{(ss)}(\omega )  \label{4f}
\end{eqnarray}%
where $A_{lij}=\frac{\left( J_{il}\right) ^{2}}{2}\left( 2-\delta
_{jl}\right) $, $B_{lij}=\frac{\left( J_{ij}\right) ^{2}+\left(
J_{jl}\right) ^{2}}{2}\left( 1-\delta _{il}\right) $ and $C_{lij}=\frac{%
\left( J_{ij}\right) ^{2}}{2}\left( 1-\delta _{jl}\right) $.

The EOMs of the correlation functions $F_{mijq}^{(X^{\left( -\right)
})}(\omega )$ and $\widetilde{F}_{imjq}^{(X^{\left( -\right) })}(\omega )$
can be directly solved, and their solutions are written as that,%
\begin{equation}
F_{mijq}^{(X^{\left( -\right) })}(\omega )=\frac{\omega ^{2}-\frac{J^{2}}{2}%
}{D_{X}\left( \omega \right) }\frac{J_{ij}J_{im}\left( 1-\delta _{mj}\right)
}{8}\left[ 3\widetilde{\chi }_{iq}(\omega )-3\chi _{jq}(\omega )\right]
\label{a5a}
\end{equation}%
\begin{equation}
\widetilde{F}_{imjq}^{(X^{\left( -\right) })}(\omega )=-\frac{\omega ^{2}-%
\frac{J^{2}}{2}}{D_{X}\left( \omega \right) }\frac{J_{ij}J_{im}\left(
1-\delta _{mj}\right) }{8}\left[ 3\chi _{iq}(\omega )-3\widetilde{\chi }%
_{jq}(\omega )\right]  \label{a5b}
\end{equation}%
where $D_{X}\left( \omega \right) =\left( \omega ^{2}-J^{2}\right) ^{2}-%
\frac{J^{4}}{4}$.

The Eqs.(\ref{a4c}-\ref{4f}) are a set of coupled equations that can be
approximately solved. As a zeroth order approximation, we first decouple
these equations by discarding the correlation functions $\widetilde{F}%
_{lijq}^{(ss)}(\omega )$, $F_{lijq}^{(\tau \tau )}(\omega )$, $\widetilde{F}%
_{lijq}^{(\tau s)}(\omega )$ and $F_{lijq}^{(s\tau )}(\omega )$ appearing in
the right hand side of these equations, in which they become independent
with each other, and we can straight solve them. Then we use these
approximation solutions to replace them that appearing in the right hand
side of other equations, respectively. Under these approximations, we
finally obtain the following solutions of the correlation functions $%
\widetilde{F}_{lijq}^{(ss)}(\omega )$, $F_{lijq}^{(\tau \tau )}(\omega )$, $%
\widetilde{F}_{lijq}^{(\tau s)}(\omega )$ and $F_{lijq}^{(s\tau )}(\omega )$%
,
\begin{equation}
\widetilde{F}_{lijq}^{(ss)}(\omega )=-\Gamma _{lij}\left( \omega \right) %
\left[ \chi _{iq}^{-+}(\omega )+\chi _{lq}^{-+}(\omega )-2\widetilde{\chi }%
_{jq}^{-+}(\omega )\right]   \label{a6a}
\end{equation}%
\begin{equation}
F_{lijq}^{(\tau \tau )}(\omega )=-\Gamma _{lij}\left( \omega \right) \left[
\widetilde{\chi }_{iq}^{-+}(\omega )+\widetilde{\chi }_{lq}^{-+}(\omega
)-2\chi _{jq}^{-+}(\omega )\right]   \label{a6b}
\end{equation}%
\begin{equation}
\widetilde{F}_{lijq}^{(\tau s)}(\omega )=-\Lambda _{lij}\left( \omega
\right) \left[ \chi _{iq}^{-+}(\omega )-\widetilde{\chi }_{lq}^{-+}(\omega )%
\right]   \label{a6c}
\end{equation}%
\begin{equation}
F_{lijq}^{(s\tau )}(\omega )=-\Lambda _{lij}\left( \omega \right) \left[
\widetilde{\chi }_{iq}^{-+}(\omega )-\chi _{lq}^{-+}(\omega )\right]
\label{a6d}
\end{equation}%
where $\Gamma _{lij}\left( \omega \right) =\frac{J_{ij}J_{jl}\left( 1-\delta
_{il}\right) }{16\left( \omega ^{2}-J^{2}\right) }\left( 1-\frac{3J^{2}}{4}%
\frac{\omega ^{2}-\frac{J^{2}}{2}}{D_{X}\left( \omega \right) }+\frac{J^{2}}{%
2\left( \omega ^{2}-\frac{J^{2}}{2}\right) }\right) $ and $\Lambda
_{lij}\left( \omega \right) =\frac{J_{ij}J_{il}\left( 1-\delta _{jl}\right)
}{16\left( \omega ^{2}-\frac{J^{2}}{2}\right) }\left( 1+\frac{J^{2}}{2\left(
\omega ^{2}-J^{2}\right) }\right) $.

Substituting the Eqs.(\ref{a5a}-\ref{a6d}) into the Eqs.(\ref{6a},\ref{6b}),
we have the solutions of the multiple-point correlation functions $%
\widetilde{F}_{ijq}^{(s)}(\omega )$ and $F_{ijq}^{(\tau )}(\omega )$,
\begin{eqnarray}
\omega \widetilde{F}_{ijq}^{(s)}(\omega ) &=&-<\widehat{\tau }_{j}^{-}%
\widehat{s}_{i}^{+}>\delta _{iq}-\left( \frac{J_{ij}}{4}+\sum_{m}\Pi
_{mij}\left( \omega \right) \right) \left[ \chi _{iq}^{-+}(\omega )-%
\widetilde{\chi }_{jq}^{-+}(\omega )\right]  \notag \\
&&+\sum_{m}\Gamma _{mij}\left( \omega \right) \left[ \chi _{iq}^{-+}(\omega
)-\chi _{mq}^{-+}(\omega )\right]  \label{a7a}
\end{eqnarray}%
\begin{eqnarray}
\omega F_{ijq}^{(\tau )}(\omega ) &=&2<\widehat{\tau }_{i}^{z}\widehat{s}%
_{j}^{z}>\delta _{jq}-\left( \frac{J_{ij}}{4}+\sum_{m}\Pi _{mij}\left(
\omega \right) \right) \left[ \widetilde{\chi }_{iq}^{-+}(\omega )-\chi
_{jq}^{-+}(\omega )\right]  \notag \\
&&+\sum_{m}\Gamma _{mij}\left( \omega \right) \left[ \widetilde{\chi }%
_{iq}^{-+}(\omega )-\widetilde{\chi }_{mq}^{-+}(\omega )\right]  \label{a7b}
\end{eqnarray}%
These solutions of the multiple-point correlation functions $\widetilde{F}%
_{ijq}^{(s)}(\omega )$ and $F_{ijq}^{(\tau )}(\omega )$ have been
incorporated the main contributions of the high order ones belonging to the $%
N=2$ level.

As a zeroth order approximation, substituting the Eqs.(\ref{6a}-\ref{6c})
into the Eqs.(\ref{5a},\ref{5b}), meanwhile discarding those related
multiple-point correlation functions belonging to the $N=2$ level, we obtain
the EOMs of the transverse and longitudinal spin susceptibilities in the
Eqs.(\ref{4a}-\ref{4c}). Substituting the Eqs.(\ref{a7a},\ref{a7b}) into the
Eq.(\ref{5a}), that including the contributions coming from the high order
related correlation functions belonging to the $N=2$ level, we obtain the
EOMs of the spin susceptibility in the Eqs.(\ref{4a1},\ref{4a2}).

\end{document}